# Improving the Limits of Detection of Low Background Alpha Emission Measurements

Brendan D. McNally [a)], Stuart Coleman, Jack T. Harris, William K. Warburton

*XIA LLC 31057 Genstar Road, Hayward, CA 94544 US*

[a)]Corresponding author: brendan@xia.com

**Abstract.** Alpha particle emission – even at extremely low levels – is a significant issue in the search for rare events (e.g., double beta decay, dark matter detection). Traditional measurement techniques require long counting times to measure low sample rates in the presence of much larger instrumental backgrounds. To address this, a commercially available instrument developed by XIA uses pulse shape analysis to discriminate alpha emissions produced by the sample from those produced by other surfaces of the instrument itself. Experience with this system has uncovered two residual sources of background: cosmogenics and radon emanation from internal components. An R&D program is underway to enhance the system and extend the pulse shape analysis technique further, so that these residual sources can be identified and rejected as well.

In this paper, we review the theory of operation and pulse shape analysis techniques used in XIA's alpha counter, and briefly explore data suggesting the origin of the residual background terms. We will then present our approach to enhance the system's ability to identify and reject these terms. Finally, we will describe a prototype system that incorporates our concepts and demonstrates their feasibility.

## INTRODUCTION

The search for rare-events, such as double beta decay or dark matter detections, is primarily a fight to reduce the incidence of background events to sufficiently low levels so that the events of interest can be seen. Two major sources of background are trace amounts of radio-impurities found within the materials used to construct, support, and shield these experiments and the contamination of these materials' surfaces caused by exposure to atmospheric radon. Characterizing surface contamination, most commonly $^{210}$Pb arising from $^{222}$Rn exposure, is a particularly challenging issue. The most commonly used screening technique, low-background counting using high purity germanium (HPGe) gamma-ray spectroscopy, is insensitive to surface contamination, since it cannot see alphas or betas and has a low efficiency for the $^{210}$Bi M x-ray at 47 keV, whose branching ratio is only 4%. Further, because this surface $^{210}$Pb does not arise from bulk isotope decay chains, sensitive HPGe measurement of the latter implies nothing about the former.

Direct detection of alpha or beta emission from $^{210}$Pb, $^{210}$Bi, and $^{210}$Po is therefore necessary to accurately characterize surface contamination. The UltraLo-1800 (UltraLo), developed by XIA, is the present state-of-the-art in commercially available low-background alpha detection, achieving sensitivity levels of 120 α/m$^2$/day in an unshielded environment at sea level. Users in such environments report achieving minimum detectable activities of surface $^{210}$Po below 0.5 mBq/m$^2$, and corresponding bulk contamination levels below 75 mBq/kg of $^{210}$Po in copper (1). Moving their UltraLo underground, users report a 5x improvement in sensitivity to bulk $^{210}$Po contamination (2), while also demonstrating that a background term is present on the surface and is likely cosmogenic in nature.

Despite improved performance when operated underground, currently available screening techniques do not have sufficient sensitivity to characterize materials at the < 1 mBq/kg (~1 α/m$^2$/day) levels required (3). To address this situation, XIA has begun to develop a new version of the UltraLo with enhanced sensitivity, thereby allowing materials to be characterized to levels approaching 1 α/m$^2$/day at the earth's surface. This enhancement is achieved by converting the UltraLo from an ionization chamber into a time projection chamber (TPC). As explained below, the

UltraLo measures certain characteristics of the ionization tracks produced by alpha particles within its volume and uses these to reject tracks that do not originate on the sample surface. Our analysis shows that residual background events are tracks either from cosmogenic events or from Rn-sourced alpha particles that mimic real events. The additional information obtained about the tracks from a TPC will therefore allow them to be rejected as well.

## THE ULTRALO-1800 LOW BACKGROUND ALPHA PARTICLE COUNTER

The UltraLo operates as a parallel plate ionization chamber with a guard ring enclosing its anode electrode (see Fig. 1). Both electrodes have charge-integrating preamplifiers, whose outputs are digitized and analyzed offline. The counter is filled with dry argon gas and operated at 1000V. Each alpha particle emitted into the chamber deposits energy in the argon, creating an ionization track. As each track electron drifts to the anode, it induces a time varying charge (current) in it that is then integrated by the preamplifier to produce an output pulse. The induction lasts exactly as long as the electron drifts, producing longer signal risetimes from longer drift distances.

Extending this analysis to the ionization track's many electrons produces distinct pulse shape characteristics based on the motion and position of the track within the counter's volume. Figure 2 shows a charge track of length $L_T$ some distance $d$ from the electrode (top), and its associated output pulse. The pulse starts at $t_0$, the instant the track forms and rises linearly as the whole track drifts across the chamber. Between time $t_S$ and $t_R$, the track's charge decreases linearly in time, as its electrons are collected, producing a parabolic preamplifier output.

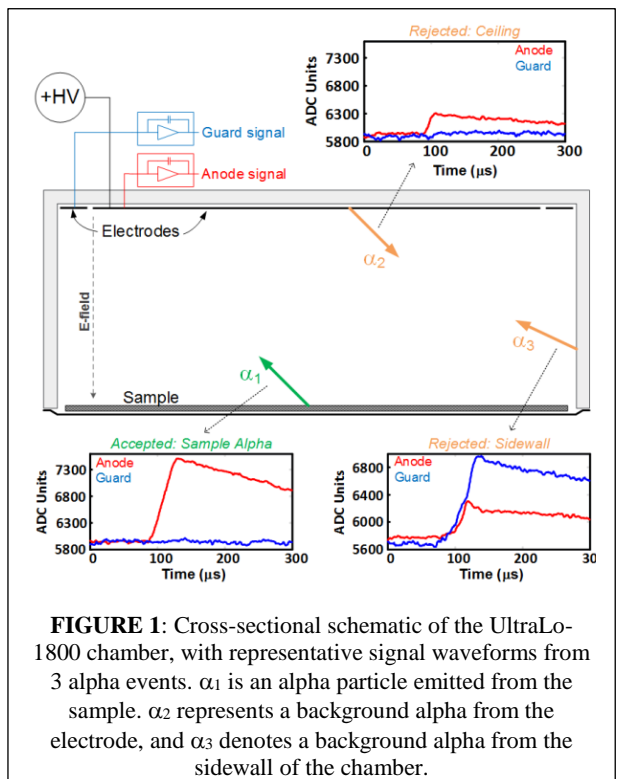

**FIGURE 1**: Cross-sectional schematic of the UltraLo-1800 chamber, with representative signal waveforms from 3 alpha events. $\alpha_1$ is an alpha particle emitted from the sample. $\alpha_2$ represents a background alpha from the electrode, and $\alpha_3$ denotes a background alpha from the sidewall of the chamber.

Thus, pulses from sample sourced alphas have a long linear risetime followed by a short parabolic section. Their total drift time $T_\alpha$ is always $t_R - t_0 = z/v_e$, a constant, where $v_e$ is the electron drift velocity. Conversely, pulses arising from alphas emitted from the electrode itself have only the parabolic section, as charges begin collecting immediately. Their maximum drift time is $L_T/v_e$ which is $\ll T_\alpha$. Typical pulses for both cases are shown in Fig. 1. Therefore, by inspecting both pulse shapes and risetimes, the UltraLo discriminates between the two cases and electronically rejects non-sample originating pulses. Because signals from alpha tracks emanating from the sidewalls have varying risetimes, we detect them directly using a guard ring to collect a significant fraction of their track charges and reject events that produce guard ring pulses. Using these methods, the UltraLo rejects emissions from all internal surfaces other than the sample tray with very high efficiency.

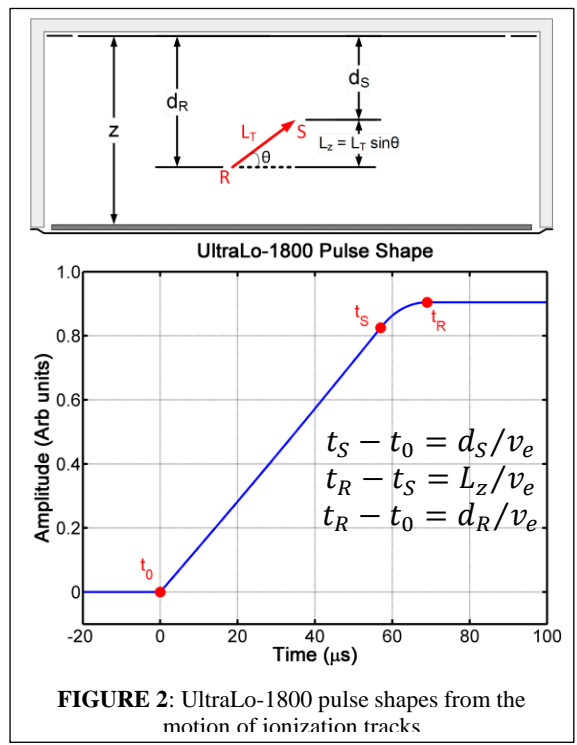

$$t_S - t_0 = d_S/v_e$$
$$t_R - t_S = L_z/v_e$$
$$t_R - t_0 = d_R/v_e$$

**FIGURE 2**: UltraLo-1800 pulse shapes from the motion of ionization tracks

# ULTRALO BACKGROUND EVENTS

Per Fig. 2, the UltraLo analysis of risetime and parabolic time provides some sensitivity to the location of charge tracks in the *z* direction. In Fig. 3, consider the radon track Rn α₁ starting at $L_R$. Its charge collection time is $L_R/v_e$ and must be $< z/v_e$. Thus by enforcing a risetime cut, the UltraLo can effectively reject most events that appear to originate in 'mid-air' (i.e., radon decays). Next consider the chamber spanning long track labeled Cosmic-1 that is produced by a minimum ionizing cosmogenic particle. This track's pulse has no linear portion and only an excessively long (equal to $z/v_e$) parabolic section. Implementing a cut on parabolic time thus allows most cosmogenic "round" events to be rejected as well. Combined, both cuts thereby create an "active" region (pink in Figure 3) in which tracks must be contained to be acceptable sample alphas.

This rejection is imperfect, however, as events Rn α₂ and Cosmic-2 in Fig. 3 show. The track from Rn α2, emitted from a radon atom in the gas and striking the sample, has an ionization track that is similar enough to α₁ to elude our cuts. Similarly, the long Cosmic-2 track's low charge density does not trigger the guard veto, yet its total charge (and *z*-projected charge) have the right location and amplitude to pass our cuts as well. Experimentally supporting this analysis, we observe related increases in both "round" events (rejected cosmogenics) and background events when our counters operate at higher elevations where there is less atmospheric overburden (4). Operating a counter underground, we have also seen radon signatures, once the larger obscuring cosmogenic term was eliminated (5).

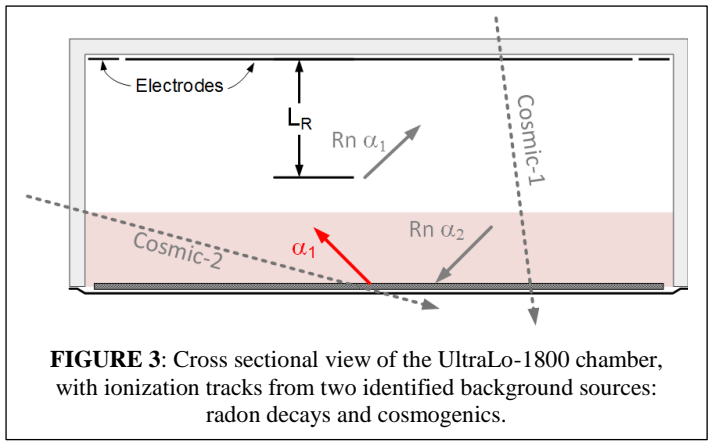

**FIGURE 3**: Cross sectional view of the UltraLo-1800 chamber, with ionization tracks from two identified background sources: radon decays and cosmogenics.

Understanding the sources of these events suggests our approach to eliminating them. First, we note that the charge tracks α₁ and Rn α₂ are not in fact identical because heavy charged particles deposit most of their energy at the end of their track, giving rise to the Bragg curve. If we can identify which end of the track contains the majority of charge, we can determine its directionality. Regarding Cosmic-2, if we could measure its projected length in *x* and *y*, as well as *z*, we could easily reject it. Rejecting these two remaining background classes thus requires more track length and charge distribution information. We propose to achieve this by converting the UltraLo into a TPC by replacing its anode and guard with a two dimensional array of small electrodes ('pixels') that are individually instrumented and read out in coincidence, where the different pixels' collected charges definitively characterize each track's length, local charge density, and orientation. In the next section, we will use a model to demonstrate the principle.

# CHARGE INDUCTION MODELING

An ionization track drifting toward a pixel array induces different amounts of charge on different pixels, from which we intend to reconstruct the original track location. In prior work (6) we developed a model that accurately describes this charge induction process and shows the pixels' output signals in time. Figure 4 shows an example (e.g., α₁). Each track's orientation is specified by *x*, *y*, and *z* of its emanation point of emanation in mm, its angle θ from the *x* axis, and its angle ϕ from the vertical. Track length is computed from the alpha's energy and its corresponding range in argon. The inset shows a sketch of a 5 x 5 section of the electrode array. The pixels are 12mm on edge, with 1mm gaps. Superimposed on the array is the *x-y* projection of the ionization track being modeled, whose width represents its local charge distribution. In these computations, the ionization track is subdivided into 50 "point charge" segments, whose charges are distributed according to the Bragg curve.

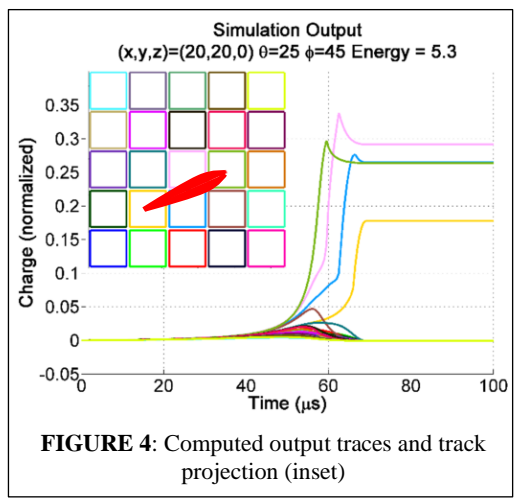

**FIGURE 4**: Computed output traces and track projection (inset)

The traces on the right are the computed preamplifier output signals, color coded by pixel. While the track forms at time $t = 0$, significant signals only appear after 45 µs, when the first charges get within 4 cm of the pixels. This is the well-known "small pixel effect" (7). Thus, the uppermost charges in the track, arrive first, under the green pixel, followed by the charges under pink, blue, and yellow pixels. The sharp rising edges on these traces make it easy to determine which parts of the track arrive first, and by extension, which end of the track is up. The peaks on some traces arise from extra induced charge that is ultimately collected on neighboring pixels. Temporarily induced charge appears on the brown and teal pixels, but no net final charge, because no part of the track intersects them. All of these features carry information about both the track orientation and charge distribution.

Figure 5 shows an identically oriented alpha track, but going in the opposite direction (e.g., Rn $\alpha_2$). Comparing to Fig. 4, the order of signal arrivals is unchanged, reflecting identical track locations in space. The final amplitudes, however, are significantly different. Compared to $\alpha_1$ the radon emission Rn $\alpha_2$ has over 30% more charge collected on the blue pixel and 15% more charge collected on the yellow pixel, and 20% less on the pink and green pixels, all due to the asymmetric nature of the Bragg curve. These readily observable differences strongly suggest that the proposed TPC design will be sufficiently sensitive to reject background radon events occurring near the sample.

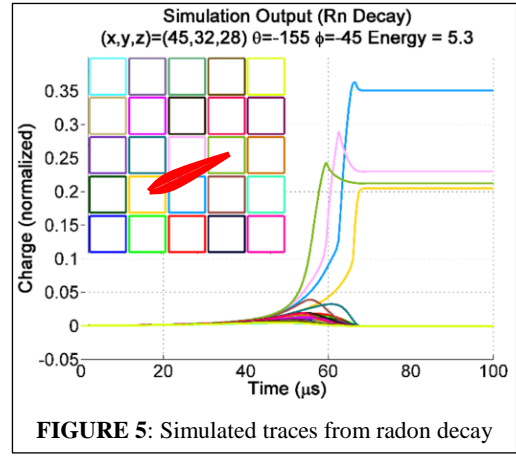

**FIGURE 5**: Simulated traces from radon decay

Figure 6 shows model results from a long ionization track like Cosmic-2. Because, above 1 MeV, protons' energy loss rate in argon is about 10X lower than that for alphas of the same energy (8), while pions' and muons' energy loss rates are yet another 5X smaller, these cosmogenics will generate long tracks with equally distributed charges throughout the counter gas. In the model, the track crosses the full detector and 6 crossed pixels collect measurable charge. Thus, a cut on the number of charge collecting pixels becomes a screen for cosmogenics. Moreover, because cosmogenic events within the "active" volume must enter the chamber through its sidewall, if we extend our pixel array to also replace the UltraLo's guard electrode we can reject cosmogenics using three criteria: 1) their low track charge density; 2) their sidewall entrance into the chamber; and, 3) their uniform charge distribution (no Bragg curve).

The large changes in detector output observed from the ionization tracks simulated above suggest that the TPC approach will be sensitive enough to allow us to easily detect and discriminate known sources of background. How effectively we can do so will depend upon what signal-to-noise we can achieve in the detector.

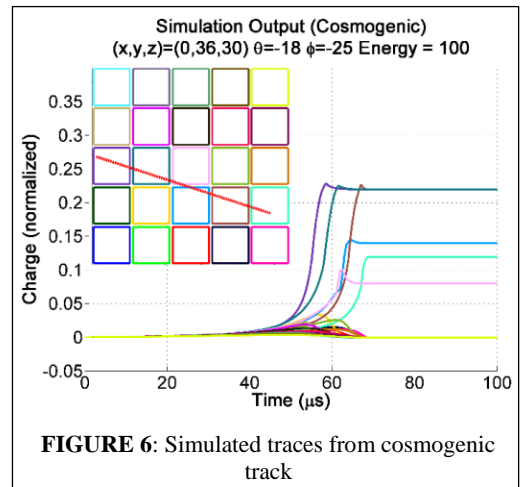

**FIGURE 6**: Simulated traces from cosmogenic track

## PROTOTYPE TPC DEVELOPMENT

We began by converting our two-channel gas ionization chamber (UltraLo) into a 64-channel small scale TPC. The core of the TPC was a new anode design comprised of an 8x8 array of square pixel elements. We modeled the effect of pixel geometry on capacitance, which impacts our anticipated noise levels, and found an optimal engineering compromise using 12mm pixels with 1mm gaps. We fabricated this electrode on a polyimide substrate and mounted it on the lower (chamber) side of chamber's ceiling plate, replacing the standard anode. As shown in Fig. 7, on the upper side of the ceiling plate we mounted an SRC card that distributed HV and connected each pixel to its corresponding preamplifier. The preamplifiers were provided 8/card, as shown. Each pixel electrode was connected to its SRC circuit by a 4-40 screw passing through the ceiling plate into a PEM nut on the SRC. Twisted pair cables were provided to both power the preamp cards and convey their signals to an external digitization and data acquisition system based on XIAs MPX-32D high-density DAQ cards on which we installed custom firmware supporting both time-synchronous waveform capture and a global triggering scheme. Later noise studies replaced the twisted pair power inputs by very low impedance bus bars, also as shown.

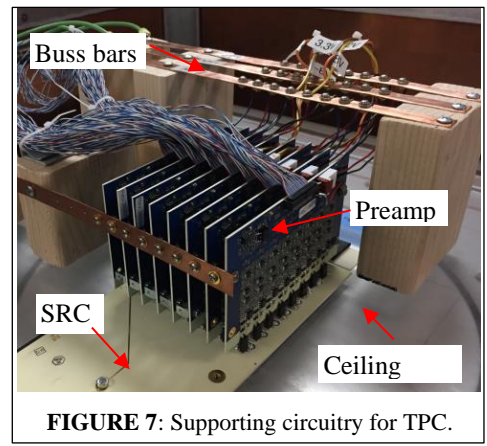

**FIGURE 7**: Supporting circuitry for TPC.

Our standard UltraLo charge-sensitive preamplifiers were designed for long term, standalone usage in industrial settings, and so were optimized for reliability, sometimes at the expense of noise performance. Recognizing the critical role noise plays in limiting our ability to analyze charge collection signals, we chose to design a new preamplifier optimized for low-noise performance. We based the new design on a topology that we had previously developed for use with ultra-low noise superconducting tunnel junction detectors (8). The design nominally achieves less than 1 nV/√Hz input referred noise. We adjusted its component values and grounding, using a well shielded test box to achieve minimum noise and then reproduced the design, 8 preamplifiers to a card using a layout that matched the 1.2 cm pixel pitch. Following resolution of further grounding and power supply noise issues, we found that we had reduced our preamplifier noise by a factor of over four, reaching a point where the noise is limited by the FET's 10pF input capacitance. Figure 8 compares the input-referred noise spectra of the original to the new design. We quote the noise improvement between 4 and 200 kHz, the bandwidth over which our signal has significant amplitude.

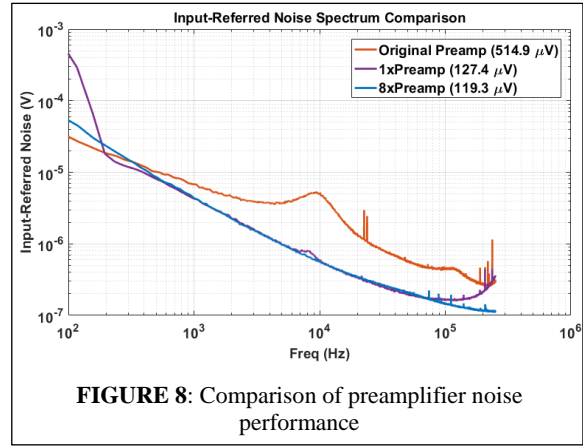

**FIGURE 8**: Comparison of preamplifier noise performance

We used 8 of these 8-channel preamplifier PCBs to instrument the 8 slots (rows) of pixel elements. Working to optimize noise performance in this configuration, we analyzed the noise in each TPC pixel channel by capturing as set of 500 baseline (no signal) waveforms and computing their average (AC) RMS noise. We then visualized the values by plotting them as a 'heatmap'. Figure 9 shows two heatmaps at different stages of noise debugging. The left image – heatmap A – is from early in the process. Each square represents an individual channel, with its color coding displaying its RMS noise in mV, per the color scale shown to the right. Channels not in operation are blue. Note that in heatmap A, the preamplifier card for slot 8 has been removed, and two preamplifiers are broken. Surprisingly, all the pixels in slot 1 (top) show extremely good noise performance of 40 µV, about 8 times lower than the average of the other slot's pixels 315 µV.

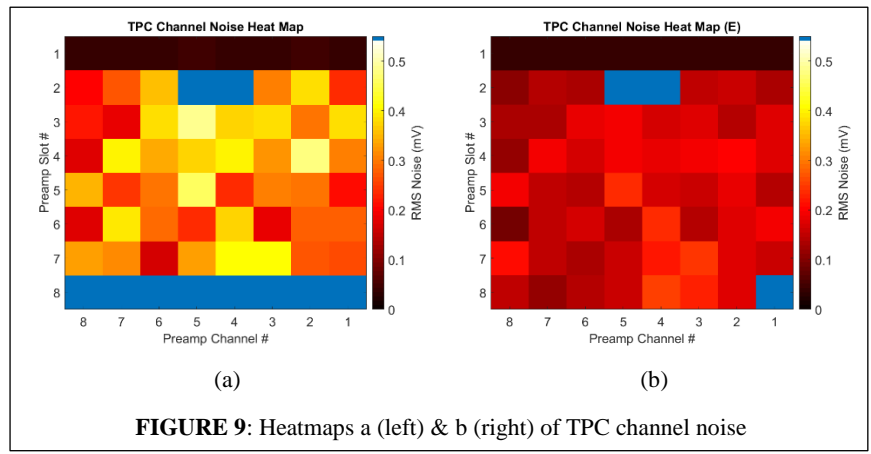

**FIGURE 9**: Heatmaps a (left) & b (right) of TPC channel noise

Using the heatmaps, we determined that noise was primarily a function of pixels physical position in the array. For example, after inserting the slot 8 preamplifier card, we found its noise was "average", eliminating the edge location as being special *per se*. Swapping preamplifier cards in slots 1 and 2 left the noise unchanged, and similarly with slots 5 and 6. Rotating the anode array by 180° in the x-y plane, however, made a major difference. The anode array itself is totally symmetric, but we were somewhat more aggressive in how tightly we tightened the pixels' connecting screws. At the conclusion of the debugging process, (Fig. 10 right) the average (RMS) noise levels had been reduce by a factor of approximately 2 to 150 µV, which is over 4X better than on the original UltraLo. Slot 1's pixels remained consistently superior, with approximately 40 µV of noise. At this point the only remaining asymmetry in the design lies in the details of the HV distribution on the SRC card.

## Data Collection and Analysis

Following noise shakedown, we placed a $^{230}$Th alpha source into the prototype TPC, and captured coincident signals from our preamplifiers. Raw data waveforms were prepared for analysis by first removing any DC offsets, and then digitally filtering with a low pass Bessel filter ($\omega_{co}$=200kHz). Figure 10 shows a typical event record after data preparation is complete. In this particular instance, one preamplifier PCB had been removed for diagnostics, so channels 56-63 were not operating, nor are they displayed. The three channels that collected charge are highlighted; channels 11, 19, and 27. For comparison to our model Fig. 4-Fig. 6, inset in Fig. 10 are the waveforms from the collecting channels 11, 19, 27, and their neighbors.

Operating on this data, our algorithm readily found amplitudes of 375, 292, and 174 in channels 11, 19, and 27, respectively. Their summed signal amplitude of 840 ADC corresponds to 4.56 MeV of alpha energy, which is close to the anticipated $^{230}$Th decay energy of 4.64 MeV. The arrival times of the 3 signals were easily distinguished and indicated that the alpha track was emitted below pixel 27 (whose signal arrived last), and ends under channel 11 (whose signal arrived first). Because more signal was collected from the track's upper end (e.g., 667 from channels 11 and 19) than from its lower end (e.g., 466 from channels 19 and 27), we infer alpha emission from the tray upward into the volume. The transient induction signal on channel 12 shows that charge collection began at 99 µs, and lasted 3.9 µs, providing an emission angle of 74° from vertical, using the known 4.56MeV alpha track length and electron drift velocity in argon. From these parameters the original track orientation was computed, as shown in Fig. 11.

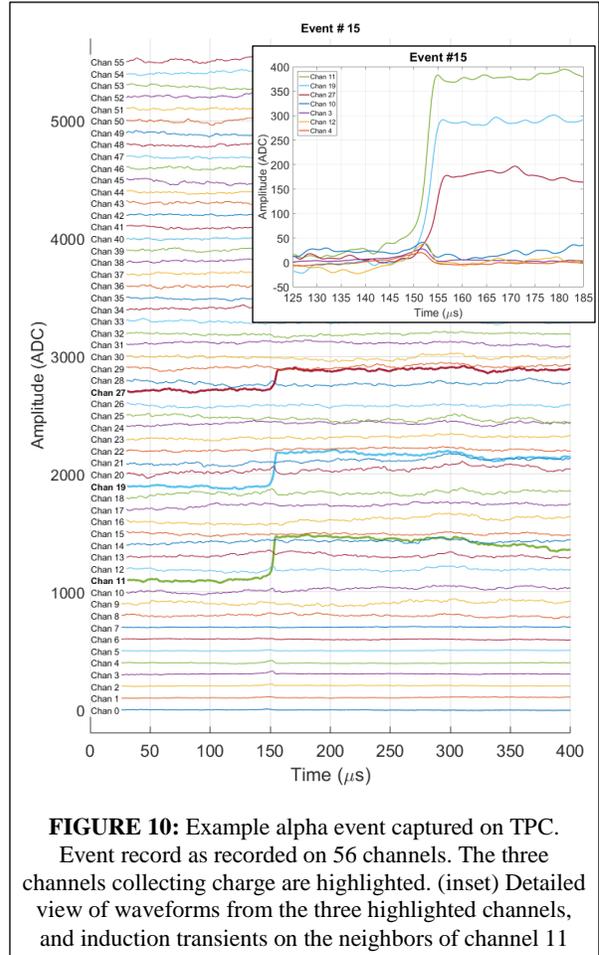

**FIGURE 10:** Example alpha event captured on TPC. Event record as recorded on 56 channels. The three channels collecting charge are highlighted. (inset) Detailed view of waveforms from the three highlighted channels, and induction transients on the neighbors of channel 11

Similarly analyzing more events in the dataset produced the resulting track orientations displayed in Fig. 12. Because our $^{230}$Th source was a 24mm diameter electrodeposited disk, in Fig. 13 (a), we overlay, for comparison, a 24mm diameter circle that contains all the emission points we extracted.

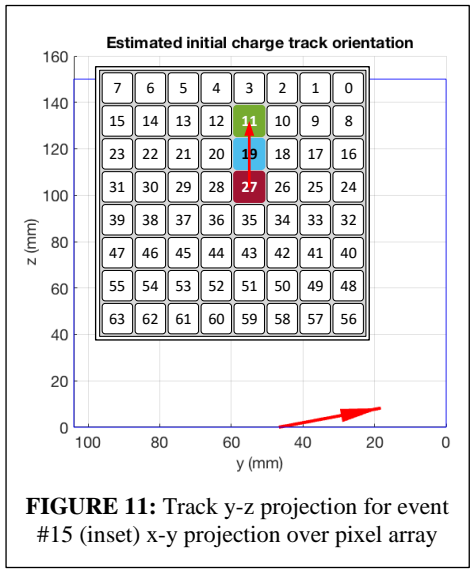

**FIGURE 11:** Track y-z projection for event #15 (inset) x-y projection over pixel array

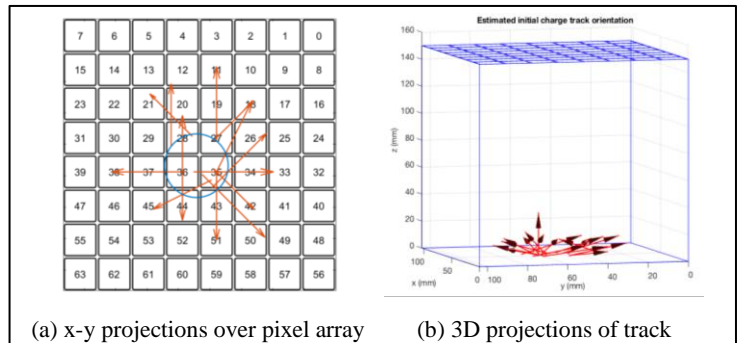

(a) x-y projections over pixel array  (b) 3D projections of track

**FIGURE 12**: track reconstructions from Th-230 dataset

This is very encouraging, since it shows that, despite the waveform noise levels in this dataset, and even with this fairly crude analysis, we are already correctly extracting the necessary parameters to estimate original track orientations with reasonable accuracy. This provides a strong proof of principle for our design concept, particularly as we expect to reduce noise levels by another factor of 4 once the full array performs as well as slot 1 currently does. Figure 13 compares noise levels in (a) the successfully analyzed example above with (b) those from an event under the pixels of slot 1. The reduction in noise levels is striking and should produce equivalent improvements in the accuracy of our pulse analysis and track reconstructions.

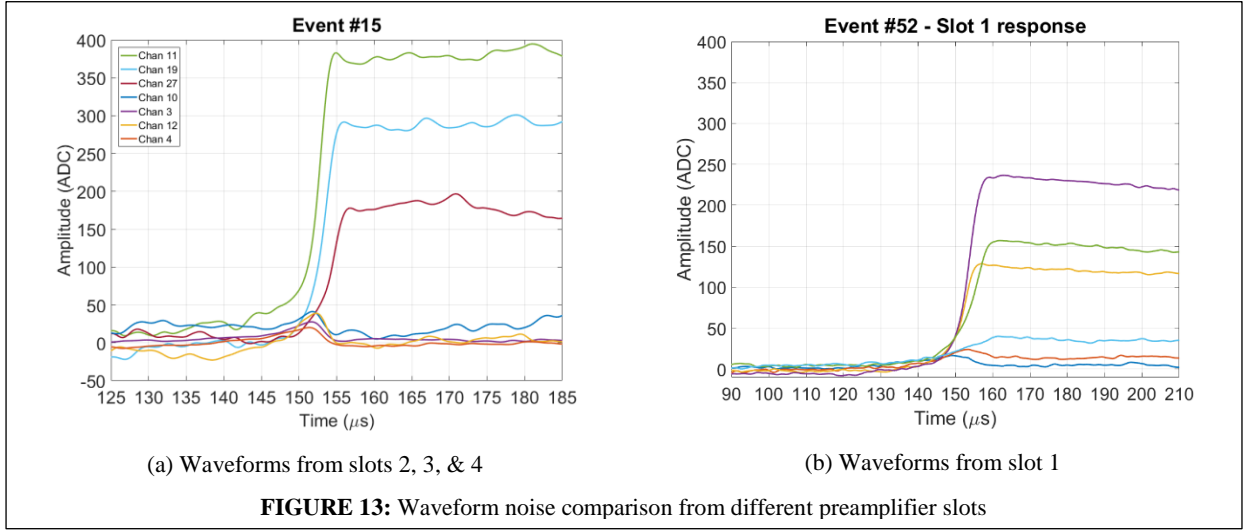

(a) Waveforms from slots 2, 3, & 4     (b) Waveforms from slot 1

**FIGURE 13:** Waveform noise comparison from different preamplifier slots

## CONCLUSIONS

We converted our two-channel gas ionization chamber (UltraLo) into a 64-channel small scale TPC. The core of the TPC was built around a new electrode design comprised of an 8x8 array of 1.3 cm square pixel elements, each of which was individually connected to a charge-sensitive preamplifier. We developed a new preamplifier design whose noise performance was improved by a factor of 4. Signals generated by an alpha source placed in the TPC were captured in coincidence and shown to match those predicted by our simulations. By analyzing these data, we demonstrated that noise levels were sufficiently low to allow emission points to be localized, the orientation of ionization tracks determined, and the distribution of charges identified. We are now ready to scale this approach up to a full-size TPC instrument, capable of achieving the desired background rates approaching 1 $\alpha/m^2/day$.

## AUTHOR CONTRIBUTIONS

B.D.M and W.K.W designed the study; B.D.M conducted the modeling; all authors assisted in designing the prototype TPC; W.K.W and J.T.H designed the preamplifier; B.D.M directed the study, and collected and analyzed the data; B.D.M and W.K.W carried out the noise reduction studies, and wrote the manuscript; W.K.W provided conceptual advice, and was a driving force behind the research.

## ACKNOWLEDGMENTS

This material is based upon work supported by the U.S. Department of Energy, Office of Science, Office of High Energy Physics, under Award Number DE-SC0015725.